\documentclass{article} 
\usepackage{iclr2027_conference,times}
\iclrfinalcopy

\usepackage{amsmath,amsfonts,bm}

\def\eqref#1{equation~\ref{#1}}

\def\1{\bm{1}}

\DeclareMathAlphabet{\mathsfit}{\encodingdefault}{\sfdefault}{m}{sl}
\SetMathAlphabet{\mathsfit}{bold}{\encodingdefault}{\sfdefault}{bx}{n}

\usepackage{hyperref}
\usepackage{graphicx}
\usepackage{tcolorbox}
\usepackage{url}
\usepackage{xspace}
\usepackage{enumitem}
\usepackage{booktabs}
\usepackage{makecell}
\usepackage{pifont}
\usepackage{wrapfig}
\usepackage{caption}
\usepackage{algorithm}
\usepackage{algorithmic}
\usepackage[table]{xcolor}
\usepackage{multirow}
\usepackage{amsmath}
\usepackage{newfloat}
\usepackage{listings}
\usepackage{xcolor}  

\def\eg{\emph{e.g.,}\xspace} 

\def\ie{\emph{i.e.,}\xspace} 

\def\method{\textbf{RepoTrace}\xspace}

\title{\method: Improving Requirement Tracing in Multi-Round Agent Interactions}

\author{
Huanyu Liu$^{1}$, Yezheng Cheng$^{2}$, Jinliang You$^{1}$, Tianxiao Huang$^{3}$, Linyu Li$^{1}$, \\ \textbf{Zheng Fang$^{1}$, Peng Cao$^{4}$, \textbf{Ge Li$^{1}$}}\thanks{Corresponding author.} \\
$^1$ Key Laboratory of High Confidence Software Technologies, Ministry of Education; \\ School of Computer Science, Peking University \quad $^2$ Xidian University \\ $^3$ School of Cyber Science and Engineering, Wuhan University $^4$ JD.com \\
\texttt{huanyuliu@stu.pku.edu.cn} \quad \texttt{lige@pku.edu.cn}
}

\begin{document}

\maketitle

\begin{abstract}

Code agents have become increasingly capable of performing software engineering tasks through interaction with software environments. However, agents tend to degrade as the number of interaction rounds increases, making it difficult to maintain existing software during iterative modification. We analyze 2,656 agent trajectories in EvoCode-Bench and identify \textbf{requirement focus misalignment} as a recurring failure pattern, where agents struggle to distinguish the primary modification targets of a requirement from unrelated code.

We propose \method, a requirement-driven repository understanding harness that combines repository structure with requirement semantics to construct a requirement-code correspondence and injects it into the agent context. This correspondence enables code agents to follow the repository architecture, trace requirements across multi-round interactions, and identify the core code elements corresponding to the current requirement. Experiments show that \method improves the mean per-round reward (MPR) across different models (\eg +19.99 on GLM-5.2) and agent harnesses (\eg +10.22 on Claude Code). Across rounds, \method increases the average per-round reward from 0.163 to 0.371 on Terminus-2 with DeepSeek-V4-Pro, while reducing average regression errors by 56.78\%. We release the source code and agents' trajectories to support future research\footnote{\url{https://github.com/gtxygyzb/RepoTrace}}.

\end{abstract}

\section{Introduction}
\label{sec:introduction}

Recently, code agents have become increasingly capable of performing software engineering tasks through interaction with software environments~\citep{claude_code, openai_codex}. A harness enables code agents to observe the environment, use tools, execute actions, receive execution feedback, and update memory~\citep{cognitive_shunyu}. Through multi-round interaction with the environment, code agents have demonstrated strong capabilities in repository-level code generation, vulnerability discovery, and software testing~\citep{anthropic2026agentic, carlini2026zerodays, maaz2026property}. Vibe coding shifts software development from one-shot code generation to iterative modification of existing software systems ~\citep{HowCodingAgentsFailTheirUsers}, where software maintenance and evolution are also important parts of the software life cycle~\citep{lehman1980programs,lientz1978maintenance}. Future software development requires agents not only to satisfy current user requirements, but also to understand the architecture of existing repositories and the potential impact of their modifications.

However, recent studies show that code agents tend to degrade as the number of interaction rounds increases in complex multi-round software engineering tasks~\citep{EvoCodeBench, Regression_Accumulation, SlopCodeBench, LLMsGetLost}. For example, EvoCode-Bench~\citep{EvoCodeBench} contains 26 stateful tasks and 227 interaction rounds from 12 mainstream code agents. The analysis of failure cases shows that the test pass rate decreases from 46.7\% in early rounds to  7.7\% in later rounds. Regression Accumulation~\citep{Regression_Accumulation} evaluates 542 tasks over 8 rounds, covering 26,016 interaction instances, and reports that 40\%–73\% of tasks lose previously satisfied behaviors during multi-round interactions.

Software has two fundamental properties: constructivity and evolvability~\citep{yang2005thinking, breivold2012software}. Constructivity refers to building software systems that satisfy functional goals based on requirements. Evolvability refers to the ability of software systems to iteratively adapt to changes in requirements, environments, and technologies while maintaining system integrity and consistency. The challenges introduced by software evolvability receive more and more attention, including architecture maintenance and dependency management~\citep{gu2026se_evolution}. Recent studies have explored self-evolving software and recursive self-improvement (RSI), further demonstrating the importance of evolvability~\citep{DarwinGodelMachine, MendelGodelMachine, HierarchicalSelf-Improvement}.

To further understand the challenges faced by code agents during multi-round interactions, we systematically analyze 2,656 agent trajectories from EvoCode-Bench. We identify \textbf{requirement focus misalignment} as a recurring failure pattern. Specifically, code agents struggle to distinguish the primary modification targets of a requirement from related code elements (\eg classes, functions, and files) that do not require direct modification. As a result, code agents may modify unrelated code or miss necessary synchronized updates. Across failure rounds, 55.16\% of code modifications occur outside the modification target files specified by the human oracle (see Section~\ref{sec:motivation_example}).



 
To this end, we propose \method, a requirement-driven repository understanding harness that improves requirement tracing for code agents in multi-round software tasks. The core idea of \method is to explicitly guide code agents to establish the correspondence between requirements and the code repository before code modification.
Specifically, \method first builds a structural representation of the code repository from code relations and computes a structural weight for each code element. The weights help identify core elements, extension points, and change propagation across the repository. Next, \method extracts the potential modification scope from the semantics of the natural-language requirement. Finally, \method combines requirement semantics with repository structure to construct a requirement-code correspondence and injects it into the agent context.
\method allows code agents to trace the repository architecture across rounds and identify the core code elements corresponding to the current requirements.


We apply \method to multiple mainstream code agents and conduct experiments on multi-round software tasks. Experimental results show that: \ding{182} Across models, \method improves the mean per-round reward (MPR) by +19.99 on GLM-5.2 and +4.91 on MiniMax-M3. \ding{183} Across harnesses, \method improves MPR by +22.25 on Terminus-2 and +10.22 on Claude Code. \ding{184} Across rounds, \method increases the average per-round reward from 0.163 to 0.371 on Terminus-2 with DeepSeek-V4-Pro, while reducing average regression errors by 56.78\%. The main contributions are summarized as follows:






\begin{enumerate}[leftmargin=*, label=\arabic*), labelsep=0.5em, itemsep=0pt, topsep=0pt]

    \item We analyze agent trajectories and identify \textbf{requirement focus misalignment} as a recurring failure pattern during multi-round interactions.

    \item We propose \method, a requirement-driven repository understanding harness that combines repository structure and requirement semantics to improve requirement tracing for agents.

    \item Experimental results show that \method improves performance across models and harnesses in multi-round interactions, and reduces regression errors.
    
\end{enumerate}

\section{Motivating Examples}
\label{sec:motivation_example}

\begin{figure}[t]
    \centering
    \vspace{-0.6cm}
    \includegraphics[width=\linewidth]{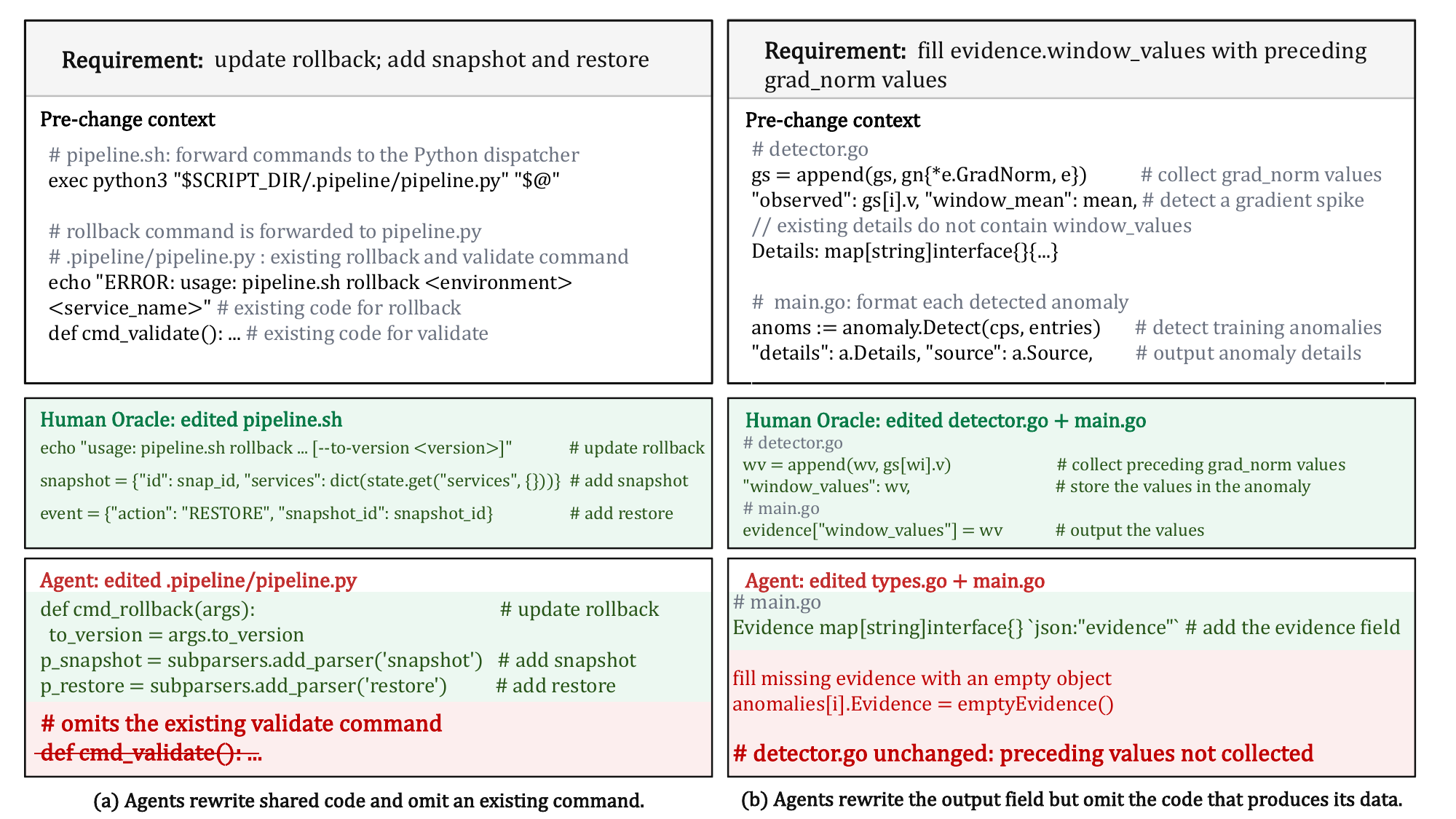}
    \vspace{-0.4cm}
    \caption{Two examples of \textbf{requirement focus misalignment}.
    (a) Unrelated code modification: the agent rewrites a shared dispatcher and omits the existing \texttt{validate()} command.
    (b) Missing update: the agent adds the output field but omits the code that collects its values.}
    \vspace{-0.4cm}
    \label{fig:motivation_example}
\end{figure}

In this section, we use two real-world examples of multi-round repository modification to illustrate \textbf{requirement focus misalignment}: code agents modify unrelated code or omit necessary updates. The examples are from real-world software development tasks in EvoCode-Bench~\citep{EvoCodeBench} and include human-written oracle solutions.

\noindent\textbf{\ding{182} Unrelated code modification.}
In Figure~\ref{fig:motivation_example}(a), the requirement asks the agent to update the \texttt{rollback} behavior and add the \texttt{snapshot()} and \texttt{restore()} functions. The human-written oracle modifies the corresponding logic in \texttt{pipeline.sh}. The agent instead rewrites \texttt{.pipeline/pipeline.py}, a shared command dispatcher that contains commands unrelated to the requirement. During the rewrite, the agent omits the existing \texttt{validate()} command and breaks the validation functionality.

\noindent\textbf{\ding{183} Missing update.}
In Figure~\ref{fig:motivation_example}(b), the requirement asks each detected anomaly to include an \texttt{evidence} object. For a gradient spike, \texttt{evidence.window\_values} records up to five preceding gradient-norm values. The values are collected in \texttt{detector.go} and formatted for output in \texttt{main.go}. The human-written oracle updates both components. However, the agent modifies only the output-related code in \texttt{types.go} and \texttt{main.go} and omits the update to \texttt{detector.go}. Consequently, the output contains the \texttt{evidence} field, but \texttt{window\_values} remains empty.

The two examples show two forms of \textbf{requirement focus misalignment}: unrelated code modifications and missing updates. Across failed interaction rounds after Round~1, 55.16\% of source-file modifications occur outside the modification target files specified by the human-written oracle (see Appendix~\ref{app:misalignment_setting} for the experimental details). The finding motivates \method, which helps code agents identify the code elements corresponding to the current requirement.

\section{Related Work}
\label{sec:related_work}




\subsection{Code Agents in Multi-Round Interactions}
Code agents have demonstrated strong capabilities in software engineering tasks. Commercial products such as Claude Code, Codex and Kimi Code~\citep{claude_code,openai_codex,moonshot2026kimicode} bring agentic coding into daily development practice. From function-level HumanEval~\citep{humaneval}, to repository-level SWE-bench~\citep{openai2024swebenchverified,SWEBenchPro}, to long-horizon tasks like Terminal-Bench~\citep{terminus2_terminal_bench}, code agents continuously achieve high scores. Many existing benchmarks focus on single-round, one-shot code generation, where software maintenance and evolution remain important throughout the software life cycle~\citep{lehman1980programs,lientz1978maintenance}. 
However, recent studies show that code agents tend to degrade as the number of interaction rounds increases.
\citet{LLMsGetLost} finds that top-tier models lose roughly 39\% of their average success rate compared to single-round settings, and \citet{Benchmarking_Correctness} reports a 20\%-27\% average drop in correct-and-secure outputs across 32 models.
Similarly, SlopCodeBench reports that no agent solves all tasks, where the best agent passes only 14.8\% of checkpoints, and 77\% of trajectories exhibit structural erosion, while 75.5\% suffer from verbosity~\citep{SlopCodeBench}.
Complementary studies report consistent findings, including declining accuracy under iterative refinement, low completion rates under stepwise requirement refinement, instruction forgetting, and progressive faithfulness loss to emergent specifications~\citep{CodeFlowBench,SR-Eval,CodeIF-Bench,SLUMP}.




\subsection{Requirements Tracing and Repository-Level Code Understanding}

Requirements traceability refers to the ability to describe and follow the life of a requirement~\citep{gotel1994analysis}, which is an active research topic in software engineering.
Recent studies explore the combination of LLMs with structured methods for requirements tracing.
TraceDev proposes a multi-agent framework that links requirements, design models, and code in a heterogeneous traceability graph to explicitly maintain consistency among software artifacts~\citep{TraceDev}.
RAG-based methods further recover requirement trace links in a largely domain-agnostic manner~\citep{REFSQ,LiSSA}.
Existing studies demonstrate the potential of language models in requirements tracing and the impact analysis of change~\citep{LLM-Driven_Cost-Effective_Req_Change_Impact_Analy}. However, they don't consider how requirement-artifact consistency should be maintained as requirements evolve over multiple rounds of development.

Meanwhile, repository-level code understanding is another key research direction.
GraphCoder proposes the code context graph (CCG), which structures information such as control flow and data dependencies into a graph to improve context retrieval for repository-level code completion~\citep{GraphCoder}. CodexGraph and RANGER organize repositories as knowledge graphs, supporting graph queries and natural-language retrieval~\citep{CodexGraph,RANGER}.
Most recently, LocAgent and DyRetriever allow agents to navigate or perform multi-hop reasoning along dependency graphs, achieving low-cost code localization and context retrieval~\citep{LocAgent,Effective_And_Efficient_Context_Retrieval}.
Existing methods enhance the understanding of code structure from different angles~\citep{RPG}. 
However, they primarily target static repositories or one-shot generation, without considering how repository structure should be maintained and updated during multi-round interactions.

\section{RepoTrace}
\label{sec:RepoTrace}

We propose \method, a requirement-driven repository understanding harness for code agents in multi-round interactions. To address requirement focus misalignment, \method{} combines repository structure with requirement semantics before code modification. The agent receives the current requirement together with structural dependencies and information about code elements associated with prior requirements.

\begin{figure}[t]
    \centering
    \vspace{-0.1cm}
    \includegraphics[width=0.97\linewidth]{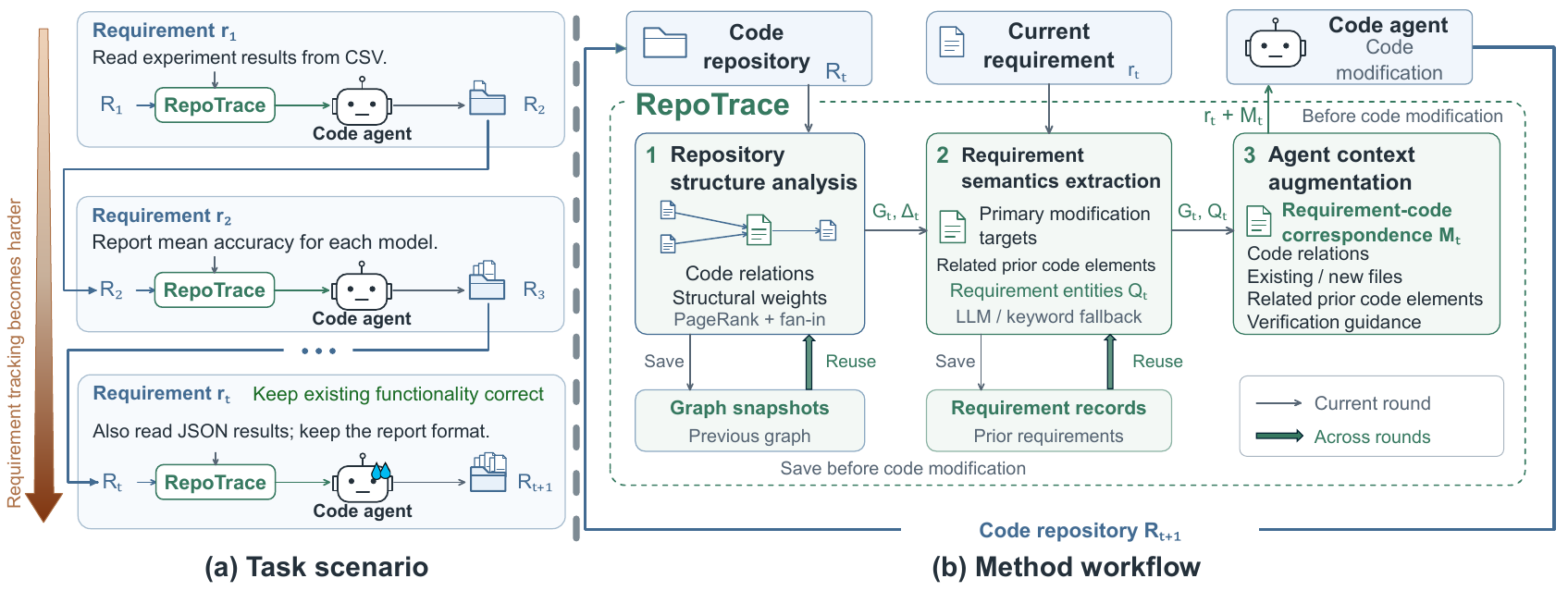}
    \vspace{-0.2cm}
    \caption{Overview of \method{} in multi-round interactions. Left: an illustrative sequence of requirements for the same code repository. Right: before each code modification, \method{} analyzes repository structure, extracts requirement semantics, and constructs a requirement-code correspondence message $M_t$ for the code agent.}
    \vspace{-0.3cm}
    \label{fig:repotrace_overview}
\end{figure}

Figure~\ref{fig:repotrace_overview} illustrates the framework. At round $t$, \method{} takes the current code repository $R_t$ and the natural-language requirement $r_t$. The following subsections introduce three steps: \ding{182} how \method{} constructs a structural representation $G_t$ of the repository; \ding{183} how \method{} extracts requirement entities $Q_t$ using the current requirement, the structural representation $G_t$, and prior requirement records; and \ding{184} how \method{} combines $G_t$ and $Q_t$ to construct a requirement-code correspondence message $M_t$, maintain requirement tracing across rounds.


\subsection{Repository Structure Analysis}
\label{sec:repotrace_structure}


\paragraph{Structural representation.}
Let $V_t$ denote the set of source files collected from $R_t$, and let $E_t$ denote the collected directed file relations. A relation $(u,v)\in E_t$ indicates that file $u$ refers to code in file $v$. The representation $G_t$ contains these file and relation sets, an index that associates symbols with their defining files, and file-level structural weights. To obtain $G_t$, we first extract the relations and symbol index, then compute the structural weights. We also compare $G_t$ with the previous representation to identify structural changes across rounds.

\paragraph{Code relations.}
To obtain the file relations and symbol index, \method{} collects source-file paths, import declarations, symbol definitions, and candidate call sites from $R_t$. Specifically, we record $(u,v)$ when file $u$ imports code from file $v$ or contains a candidate call site whose name matches a symbol defined in $v$. We resolve repository-local imports and aggregate matched calls into file relations. These relations provide an approximate structural view, and the symbol index connects names in a requirement to repository files.

\paragraph{Structural weights.}
To identify core files that may affect dependent files, we compute two structural scores from code relations. We apply PageRank~\citep{Page1999ThePC} to the file graph formed by nodes $V_t$ and relations from $E_t$ whose endpoints are both in $V_t$. We also compute the fan-in of each file, defined as the number of distinct files that refer to it. PageRank captures the structural importance of files in the dependency graph, while fan-in measures how many files directly depend on each file. We use PageRank to prioritize files included in the LLM prompt and to order examples of dependent files. We use fan-in to select files referenced by multiple files and to rank the existing-file overview in the agent context. Two scores expose structural dependencies \textbf{without} assuming that structurally important files require modification.


\paragraph{Structural changes.}
To reflect code modifications across rounds, \method{} reconstructs $G_t$ from the current repository before processing each requirement. To identify what has changed, we compare $G_t$ with the most recent retained representation $G_{\mathrm{prev}}$, whose file and relation sets are $V_{\mathrm{prev}}$ and $E_{\mathrm{prev}}$. We summarize structural changes as
\begin{equation}
\Delta_t =
\bigl(
V_t \setminus V_{\mathrm{prev}},\;
V_{\mathrm{prev}} \setminus V_t,\;
E_t \setminus E_{\mathrm{prev}},\;
E_{\mathrm{prev}} \setminus E_t
\bigr).
\label{eq:repotrace_delta}
\end{equation}
The four components record added files, removed files, added relations, and removed relations, respectively. For the first round, the previous file and relation sets are empty. This summary provides structural changes observed before the current modification, including changes made during preceding rounds. The analysis produces the current structural representation $G_t$ and the change summary $\Delta_t$ for requirement semantics extraction. For specific examples, see Appendix~\ref{app:repotrace_structure}.

\subsection{Requirement Semantics Extraction}
\label{sec:repotrace_semantics}
Building on $G_t$ and $\Delta_t$, \method{} analyzes the current requirement $r_t$ and prior requirements $r_{<t}$, in order to identify candidate primary modification targets and potentially affected code elements.
To connect $r_t$ with $r_{<t}$, \method{} retains a history $\mathcal{H}_{<t}$ of earlier requirement records, including their extracted code elements and summaries. \method{} combines $r_t$, $G_t$, $\Delta_t$, and $\mathcal{H}_{<t}$ to produce a structured record of requirement entities:
\begin{equation}
Q_t = \operatorname{Extract}
\bigl(r_t, G_t, \Delta_t, \mathcal{H}_{<t}\bigr).
\label{eq:repotrace_extraction}
\end{equation}
The record $Q_t$ contains candidate target files and symbols, behavior constraints, potentially affected prior code elements, and a requirement summary. These fields represent two roles: candidate primary modification targets and potentially affected prior code elements.

\ding{182} For LLM-based extraction, \method{} calls an LLM to analyze the current requirement, a list of files ranked by PageRank, symbol-to-file associations, and a bounded summary of recent requirement records. When prior records exist, \method{} also includes a compact summary of $\Delta_t$ in the prompt. The LLM extracts the requirement entities and returns the structured record $Q_t$. \ding{183} If no LLM is configured or extraction fails, a keyword fallback matches file paths and identifiers in $r_t$ against the repository index. The fallback compares these matches with historical targets to distinguish newly identified targets from potentially affected prior code elements without using $\Delta_t$. In both paths, the records retain associations between requirements and code elements across rounds. For specific examples, see Appendix~\ref{app:repotrace_prompt}.


\subsection{Agent Context Augmentation across Rounds}
\label{sec:repotrace_tracking}

The requirement semantics extraction produces $Q_t$, which identifies candidate primary modification targets and potentially affected prior code elements. \method{} combines this requirement-specific information with the structural representation $G_t$ to construct a requirement-code correspondence message for the code agent. To maintain requirement tracing across rounds, \method{} also retains the current structural representation and requirement record for use in subsequent rounds.

\paragraph{Requirement-code correspondence.}
The structural representation $G_t$ describes code relations, while $Q_t$ identifies candidate code elements associated with the current and prior requirements. To provide the code agent with context for the current requirement, we select information from both to construct the requirement-code correspondence message:
\begin{equation}
M_t = \operatorname{Compose}(G_t,Q_t),
\qquad
\widetilde{r}_t = r_t \circ M_t,
\label{eq:repotrace_context}
\end{equation}
where $\circ$ denotes text concatenation. The message combines four types of information: code relations for files with high fan-in and selected direct dependents; an overview of existing files and candidate new files; potentially affected prior code elements; and verification guidance. We identify candidate new files by comparing the target paths in $Q_t$ with $V_t$. The verification guidance asks the agent to build and test the modified repository and suggests a command based on the programming languages in the repository. The message uses selected fields of $Q_t$, rather than reproducing the entire extraction record. The code agent receives the original requirement with this supplemental context and retains responsibility for code modification and verification.

\paragraph{Maintaining records across rounds.}
To reuse the current analysis in later rounds, \method{} stores the current graph representation and appends the current requirement record to its history after constructing $M_t$ and before the code agent executes. The agent then modifies the repository, whose state is observed as $R_{t+1}$ at the next round. At that round, \method{} reconstructs the repository structure, compares it with the retained graph, and reuses prior requirement records during extraction, repeating the three steps for the next requirement. Graph snapshots and requirement records therefore describe prior analyses, rather than verified completion states. For specific examples, see Appendix~\ref{app:repotrace_message}.

\section{Experiments}
\label{sec:experiments}

To evaluate the effectiveness of \method, we conduct experiments with multiple code agents on multi-round software tasks. In this section, we introduce the benchmarks, baselines, models, and evaluation metrics. We then present the experimental results and analysis.

\subsection{Experimental Setup}
\label{subsec:exp_setup}

\paragraph{Benchmarks.}
We evaluate code agents on EvoCode-Bench~\citep{EvoCodeBench}, which contains 26 stateful software engineering tasks and 227 evaluated rounds across six technical domains, including ML/MLOps, data engineering, systems and code, scientific computing, testing and automation, and infrastructure and security. Each task contains 5-15 rounds, with the same workspace and agent session preserved across rounds. Each round introduces a new requirement and uses cumulative test cases to verify the current and previous active requirements. A round receives a reward only when all test cases pass.

\paragraph{Baselines and Models.}
We integrate \method into two agent harnesses, Terminus-2~\citep{laude2026terminus2} and Claude Code~\citep{claude_code}, within the Harbor agent environment~\citep{laude2026harbor}. We evaluate four models: GLM-5.2~\citep{glm}, DeepSeek-V4-Pro~\citep{deepseekv4}, MiniMax-M3~\citep{lai2026minimaxsparseattention}, and Qwen3.7-Max~\citep{qwen37}. We access all models via API. All API configurations (\eg temperature, top-$k$, and thinking effort) use the default settings.

\paragraph{Evaluation Metrics.}
We follow the evaluation setting of EvoCode-Bench~\citep{EvoCodeBench}.
\begin{enumerate}[label=\arabic*), leftmargin=*, nosep]
\item \textbf{Mean per-round reward (MPR) $\uparrow$} measures the fraction of passed rounds, where a round is counted as passed only when all test cases pass. The reward is averaged across all tasks.

\item \textbf{Final-round completion rate (FCR) $\uparrow$} measures the average test case completion rate of the final round across all tasks. The test cases are cumulative, so completing the final-round test cases means that all requirements in the requirement chain are satisfied.

\item \textbf{Perfect-task rate (PTR) $\uparrow$} measures the fraction of tasks in which a single attempt passes every round.

\item \textbf{Hard-task rate (HTR) $\downarrow$} measures the fraction of tasks whose per-task score is at most 0.5.

\item \textbf{Test Case Score (TCS) $\uparrow$} measures the average test case completion rate of the entire dataset.

\item \textbf{Input / Output Tokens (IT / OT) $\downarrow$} measure the total number of input / output tokens used across all rounds of all tasks in the dataset, expressed in millions of tokens (M).
\end{enumerate}

\subsection{RQ1: To what extent does \method improve performance across different models?}
\label{subsec:rq1}
\begin{table}[t]
  \caption{Main results and token usage on EvoCode-Bench.}
  \vspace{-0.1cm}
  \label{tab:evocodebench_token}
  \centering
  \small
  \resizebox{0.75\linewidth}{!}{
    \begin{tabular}{lccccccc}
      \toprule
       & MPR~$\uparrow$ & FCR~$\uparrow$ & PTR~$\uparrow$ & HTR~$\downarrow$ & TCS~$\uparrow$ & IT(M)~$\downarrow$ &OT(M)~$\downarrow$ \\
      \midrule
      DeepSeek-V4-Pro      & 14.14          & 11.54          & 3.85           & 92.31          & 61.74            & 441.33           & \textbf{12.51} \\
      \quad w/ RepoTrace    & \textbf{36.39} & \textbf{34.62} & \textbf{15.38} & \textbf{61.54} & \textbf{73.63}   & \textbf{319.48}  & 16.92          \\
      \midrule
      GLM-5.2              & 16.19          & 3.85           & 3.85           & 88.46          & 47.52            & \textbf{98.26}   & \textbf{7.62}  \\
      \quad w/ RepoTrace    & \textbf{36.18} & \textbf{30.77} & \textbf{7.69}  & \textbf{61.54} & \textbf{77.33}   & 175.62           & 12.30          \\
      \midrule
      MiniMax-M3           & 23.42          & \textbf{23.08} & \textbf{7.69}  & \textbf{73.08} & 61.46            & \textbf{726.66}  & \textbf{5.89}  \\
      \quad w/ RepoTrace    & \textbf{28.33} & 15.38          & \textbf{7.69}  & \textbf{73.08} & \textbf{76.76}   & 1445.57          & 7.99           \\
      \midrule
      Qwen3.7-Max          & \textbf{11.87} & \textbf{7.69}  & 0.00           & \textbf{92.31} & \textbf{67.37}   & \textbf{227.49}  & \textbf{7.89}  \\
      \quad w/ RepoTrace    & 11.36          & 3.85           & 0.00           & \textbf{92.31} & 66.06            & 275.64           & 9.07           \\
      \bottomrule
    \end{tabular}}
    \vspace{-0.4cm}
\end{table}

To answer RQ1, we evaluate four representative models with and without \method on EvoCode-Bench, in order to verify whether \method improves model performance on multi-round software engineering tasks. Table~\ref{tab:evocodebench_token} reports the results.

\textbf{\method brings improvement on stronger models that can effectively exploit repository-level guidance,} while the benefit is limited for models that struggle to utilize it.
On DeepSeek-V4-Pro, \method raises MPR from 14.14 to 36.39 (+22.25, about $2.57\times$), with FCR and PTR improving by about $3\times$ and $4\times$ respectively. This shows that the traced repository-level guidance translates into both higher per-round success and more fully completed tasks. The benefit is limited for Qwen3.7-Max, suggesting that the advantage scales with the model's ability to exploit the injected guidance.
The extra token cost is acceptable. On DeepSeek-V4-Pro, \method reduces IT from 441.33M to 319.48M ($-27.61\%$) while tripling MPR, since repository-level guidance prunes redundant exploration of the codebase. Considering that OT rises only moderately across all models and IT even decreases on the strongest performer, we regard the overall token overhead as acceptable.
Overall, \method improves model performance on EvoCode-Bench with an acceptable efficiency cost.

\subsection{RQ2: To what extent does \method improve performance across different harnesses?}
\label{subsec:rq2}
\begin{table}[ht]
  \vspace{-0.1cm}
  \caption{RepoTrace across agent harnesses on EvoCode-Bench.}
  \vspace{-0.15cm}
  \label{tab:evocodebench_harness}
  \centering
  \small
  \resizebox{0.70\linewidth}{!}{
    \begin{tabular}{lccccccc}
      \toprule
      Agent & MPR~$\uparrow$ & FCR~$\uparrow$ & PTR~$\uparrow$ & HTR~$\downarrow$ & TCS~$\uparrow$ & IO~$\downarrow$ & OT~$\downarrow$ \\
      \midrule
      Terminus-2          & 14.14          & 11.54          & 3.85           & 92.31          & 61.74           & 441.33          & \textbf{12.51} \\
      \quad w/ RepoTrace   & \textbf{36.39} & \textbf{34.62} & \textbf{15.38} & \textbf{61.54} & \textbf{73.63}  & \textbf{319.48} & 16.92          \\
      \midrule
      Claude Code         & 44.65          & 42.31          & 19.23          & 57.69          & 91.14           & 842.06          & \textbf{7.99}  \\
      \quad w/ RepoTrace   & \textbf{54.87} & \textbf{46.15} & \textbf{26.92} & \textbf{50.00} & \textbf{93.02} & \textbf{675.36} & 9.44           \\
      \bottomrule
    \end{tabular}}
    \vspace{-0.15cm}
\end{table}

To answer RQ2, we evaluate \method on two harnesses (\ie Terminus-2 and Claude Code, both driven by DeepSeek-V4-Pro) to verify whether the effectiveness of \method generalizes across different agent harnesses. Table~\ref{tab:evocodebench_harness} reports the results.

\textbf{\method improves performance across different harnesses.}
On the weaker Terminus-2, \method raises MPR from 14.14 to 36.39 (about $2.57\times$) and PTR from 3.85 to 15.38 (about $4\times$).
On Claude Code, \method further improves MPR from 44.65 to 54.87 (+10.22 points), suggesting that the repository-level context extracted by \method complements the context extracted by the harness (see RQ4 and Appendix~\ref{app:repotrace_examples}).
Token efficiency improves consistently as well when using DeepSeek-V4-Pro: IT decreases by $27.61\%$ on Terminus-2 and by $19.80\%$ on Claude Code,  while OT rises only slightly on both harnesses.
Overall, \method brings consistent improvement across harnesses at an acceptable efficiency cost.

\subsection{RQ3: How does \method's improvement vary across rounds?}
\label{subsec:rq3}


\begin{figure}[t]
    \centering
    \begin{minipage}[b]{0.49\linewidth}
        \centering
        \includegraphics[width=\linewidth]{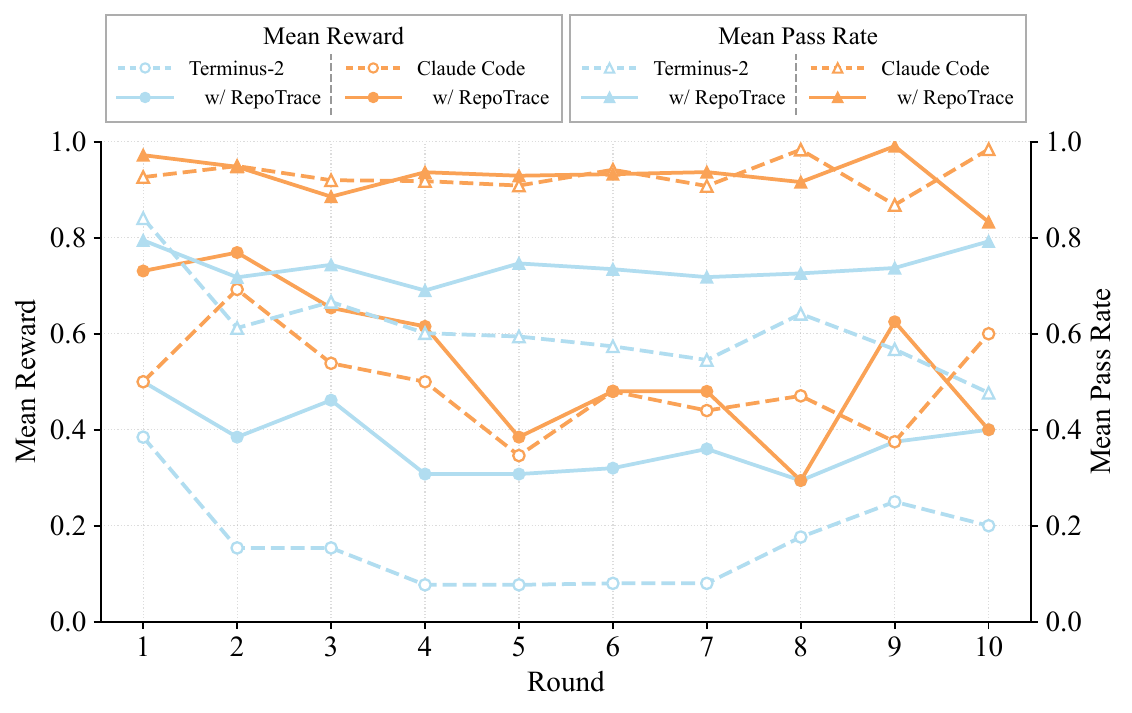}
        \\[1mm]
        \vspace{-0.2cm}
        {\small (a) Mean Reward and Mean Pass Rate}
    \end{minipage}%
    \hfill
    \begin{minipage}[b]{0.49\linewidth}
        \centering
        \includegraphics[width=\linewidth]{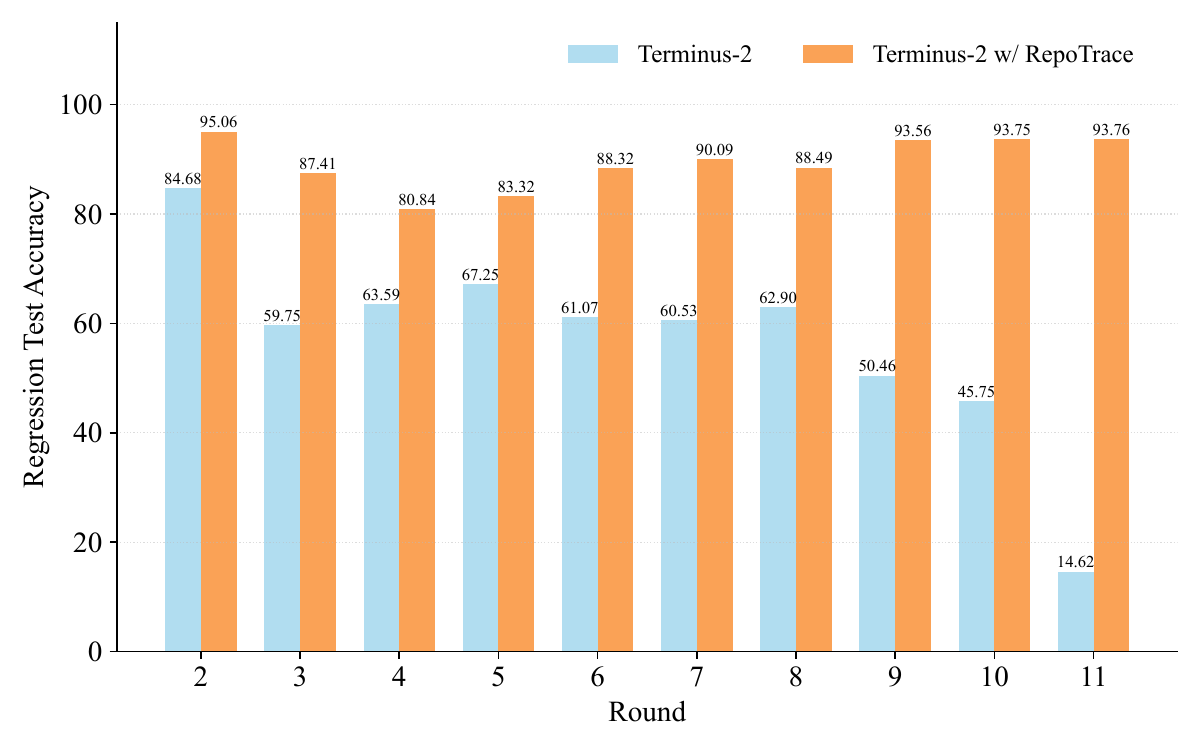}
        \\[1mm]
        \vspace{-0.2cm}
        {\small (b) Regression Test Accuracy}
    \end{minipage}
    \caption{Results of DeepSeek-V4-Pro on EvoCode-Bench using Terminus-2 and Claude Code, with and without \method. \textbf{(a)} Mean reward (left axis) and mean pass rate (right axis) over 10 rounds: the average reward and the average  fraction of test cases passed per round, each averaged over tasks. \textbf{(b)} Regression test accuracy in multi-round interactions: for each round $i \in \{2,\ldots,11\}$, we report the mean pass rate over all test cases belonging to rounds $1,\ldots,i-1$, averaged over 26 tasks. Higher values indicate fewer regression errors.}
    \label{fig:round_curve}
\end{figure}


\begin{figure}[t]
    \centering
    \vspace{-0.25cm}
    \includegraphics[width=0.95\linewidth]{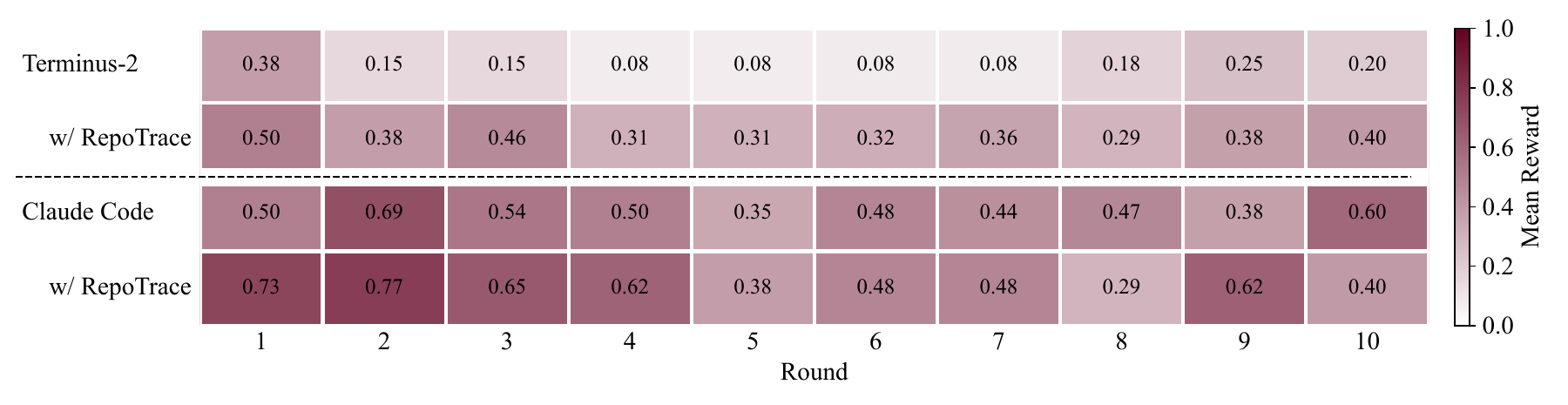}
    \vspace{-0.35cm}
    \caption{Heatmap of the per-round mean reward for each harness-\method combination.}
    \vspace{-0.4cm}
    \label{fig:reward_heatmap}
\end{figure}


To answer RQ3, we analyze the per-round behavior of DeepSeek-V4-Pro on EvoCode-Bench under both harnesses over 10 rounds. 
Figure~\ref{fig:round_curve} reports the per-round mean reward, mean pass ratio, and the regression errors on earlier rounds' test cases, and Figure~\ref{fig:reward_heatmap} visualizes the per-round mean reward of each harness–\method combination.

\textbf{\method mitigates error accumulation and regression across rounds.}
On Terminus-2, vanilla collapses after the first round, dropping from 0.38 in round 1 to 0.08 $\sim$ 0.15 in rounds 2 to 7. In contrast, \method maintains a mean reward between 0.29 and 0.50 and leads in all 10 rounds. The regression results further confirm this gap. Vanilla's pass rate on the test cases of earlier rounds degrades from 84.68 at round 2 to only 14.62 at round 11, while \method keeps this rate above 80 in every round. Averaged over rounds 2 to 11, \method improves this pass rate from 57.06 to 89.46 (+32.40 points, a 56.78\% relative improvement), which means fewer regression errors.
Overall, the benefit of \method persists as rounds advance, with a steady improvement in both harnesses. 

\subsection{RQ4: How Does \method Improve Requirement Tracing?}
\label{subsec:rq4}

\begin{figure}[t]
\centering
\vspace{-0.1cm}
\includegraphics[width=1.0\linewidth]{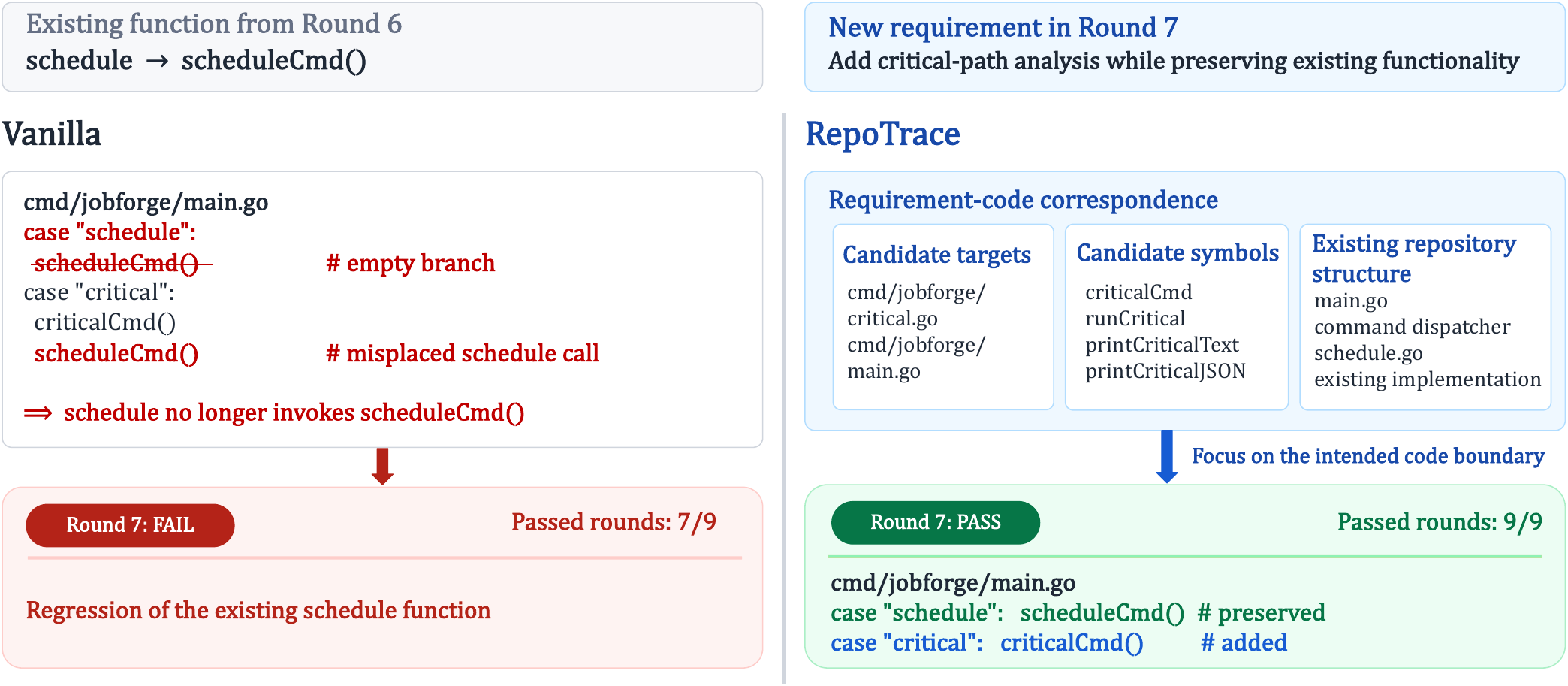}
\caption{Case study of \method on JobForge. In the Vanilla setting, adding the \texttt{critical} subcommand breaks the existing \texttt{schedule} dispatch. In the \method setting, the requirement-code correspondence guides the agent to add an independent \texttt{critical} branch while preserving the \texttt{schedule} function, satisfying both the current and prior requirements.}
\vspace{-0.3cm}
\label{fig:RQ4_Case_Study}
\end{figure}

In RQ4, we use a case study to further examine how \method improves requirement tracing. JobForge is a nine-round software task from EvoCode-Bench. A code agent incrementally develops a Go command-line tool as new requirements arrive. Round~6 adds the \texttt{schedule} subcommand. Round~7 adds the \texttt{critical} subcommand for critical-path analysis while requiring existing functionality to remain correct. We compare Vanilla and \method trajectories produced with the same requirement chain, model, and execution environment. We independently replayed both trajectories and confirmed their outcomes. Figure~\ref{fig:RQ4_Case_Study} compares the code modifications under the two settings.

\textbf{\method aligns code modifications focused on the current requirement and reduces unintended changes to existing functionality.}
\ding{182} In the Vanilla, the agent added \texttt{critical} to the shared command dispatcher but left the existing \texttt{schedule} branch without a call to \texttt{scheduleCmd()}. The command dispatcher routes each subcommand to its implementation. The missing call therefore broke the scheduling functionality introduced in Round~6. \ding{183} In the \method, the requirement-code correspondence identified \texttt{cmd/jobforge/critical.go} and \texttt{cmd/jobforge/main.go} as candidate primary modification targets, and \texttt{criticalCmd} and related critical-path functions as candidate symbols. The repository structure also showed the existing \texttt{schedule} implementation. Using the correspondence, the agent added a separate dispatch branch for \texttt{critical} and preserved the existing \texttt{schedule}~$\rightarrow$~\texttt{scheduleCmd()} relation. Overall, \method combines requirement semantics with repository structure to identify primary modification targets and preserve existing functionality, reducing requirement focus misalignment.

\section{Discussion}
\label{sec:discussion}

In this section, we discuss 1) the applicability of \method and 2) the role of requirements in agentic software engineering.

As a requirement-driven repository understanding method, \method can be extended to different harnesses through the requirement-code correspondence produced by \method. Both humans and code agents can modify the correspondence, allowing \method to work with different harnesses without directly modifying their execution process. Deeper integration (\eg using hooks) may provide more control but reduce the range of applicable \method. Supporting a wider range of harnesses while maintaining effective integration remains an open problem.

Requirements play a central role in agentic software engineering. Code agents need to translate requirements into code changes and retain previous requirements as the software evolves. As the interaction horizon grows, agents may get lost, focus on the most recent requirement, and miss requirements addressed in earlier rounds. Therefore, maintaining requirements throughout software development can help agents preserve their objectives while making new changes. More broadly, requirements provide a persistent source of objectives for agents to build on their previous experience, which is important for RSI.




\section{Conclusion and Future Work}
\label{sec:conclusion}

We present \method, a requirement-driven repository understanding harness that improves requirement tracing in multi-round agent interactions. \method helps code agents focus on relevant code and preserve existing functionality across interactions. In future work, we plan to: 1) explore how to manage and abstract requirement changes through structured representations, as with code, and 2) study how \method can help promote RSI.

\section*{AI use statement}
In preparing this manuscript, we use LLMs to assist with writing and coding. Specifically, LLMs improve clarity, grammar, and phrasing, ensuring the text is concise and readable. The use of LLMs does not influence the technical contributions or the interpretation of experimental findings. All content edited by LLMs was carefully checked by at least two authors.

\bibliography{ref}
\bibliographystyle{iclr2027_conference}

\newpage

\appendix

\section{Specific Examples of RepoTrace}
\label{app:repotrace_examples}

We use JobForge Round~7 to illustrate the three steps in Section~4: repository structure analysis, requirement semantics extraction, and agent context augmentation across rounds. JobForge is a Go command-line tool developed over multiple rounds. Round~6 introduces the \texttt{schedule} subcommand. In Round~7, the requirement $r_7$ asks for a \texttt{critical} subcommand that reports the longest dependency chain by job count and preserves all existing behavior. The following examples connect this requirement to the structural representation $G_7$, requirement entities $Q_7$, and requirement-code correspondence message $M_7$.

\subsection{Repository Structure Analysis}
\label{app:repotrace_structure}
\begin{figure}[!ht]
    \centering
    \includegraphics[width=\linewidth]{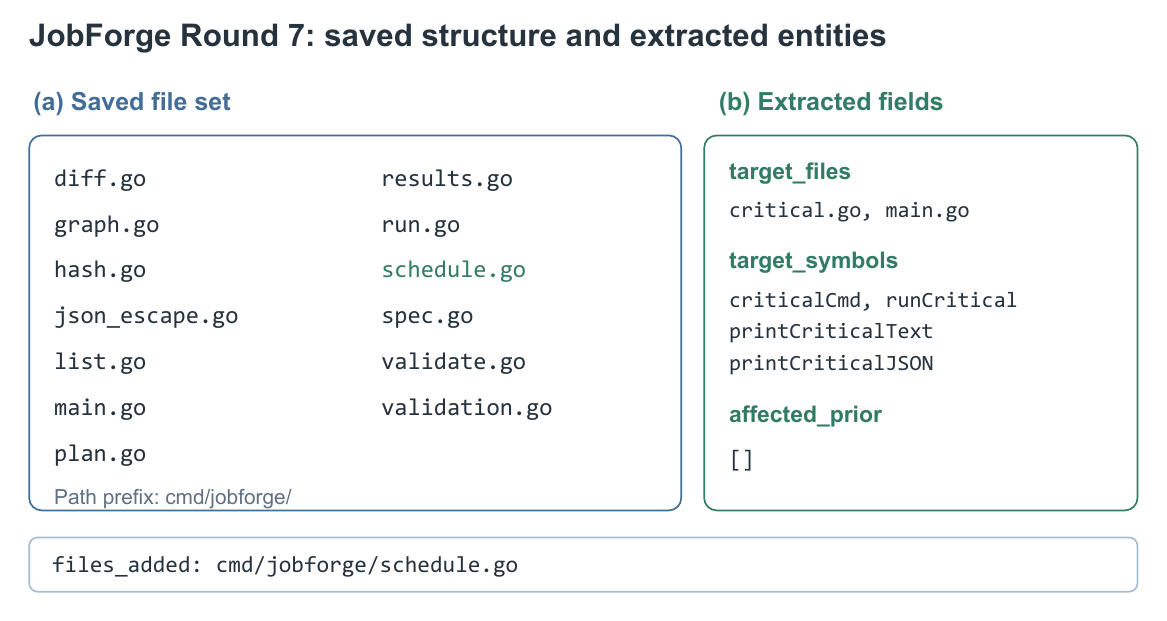}
    \caption{Repository structure and selected requirement entities for JobForge Round~7. Panel (a) shows the source-file set $V_7$, and panel (b) shows candidate targets and potentially affected prior code elements from $Q_7$. The bottom row shows the added-file entry in $\Delta_7$. These records distinguish the existing \texttt{schedule.go} file from the candidate new file \texttt{critical.go}.}
    \label{fig:repotrace_structure_example}
\end{figure}

Following Section~4.1, the structural representation describes the repository before the current code modification. Figure~\ref{fig:repotrace_structure_example}(a) lists the 13 source files in $V_7$, including \texttt{main.go} and \texttt{schedule.go}. All listed paths share the prefix \texttt{cmd/jobforge/}. The structural change summary $\Delta_7$ records \texttt{schedule.go} as an added file relative to the preceding snapshot. This entry captures a change observed before Round~7, rather than a file proposed for the current requirement.

\subsection{Requirement Semantics Extraction}
\label{app:repotrace_prompt}
\begin{figure}[!ht]
    \centering
    \includegraphics[width=0.78\linewidth]{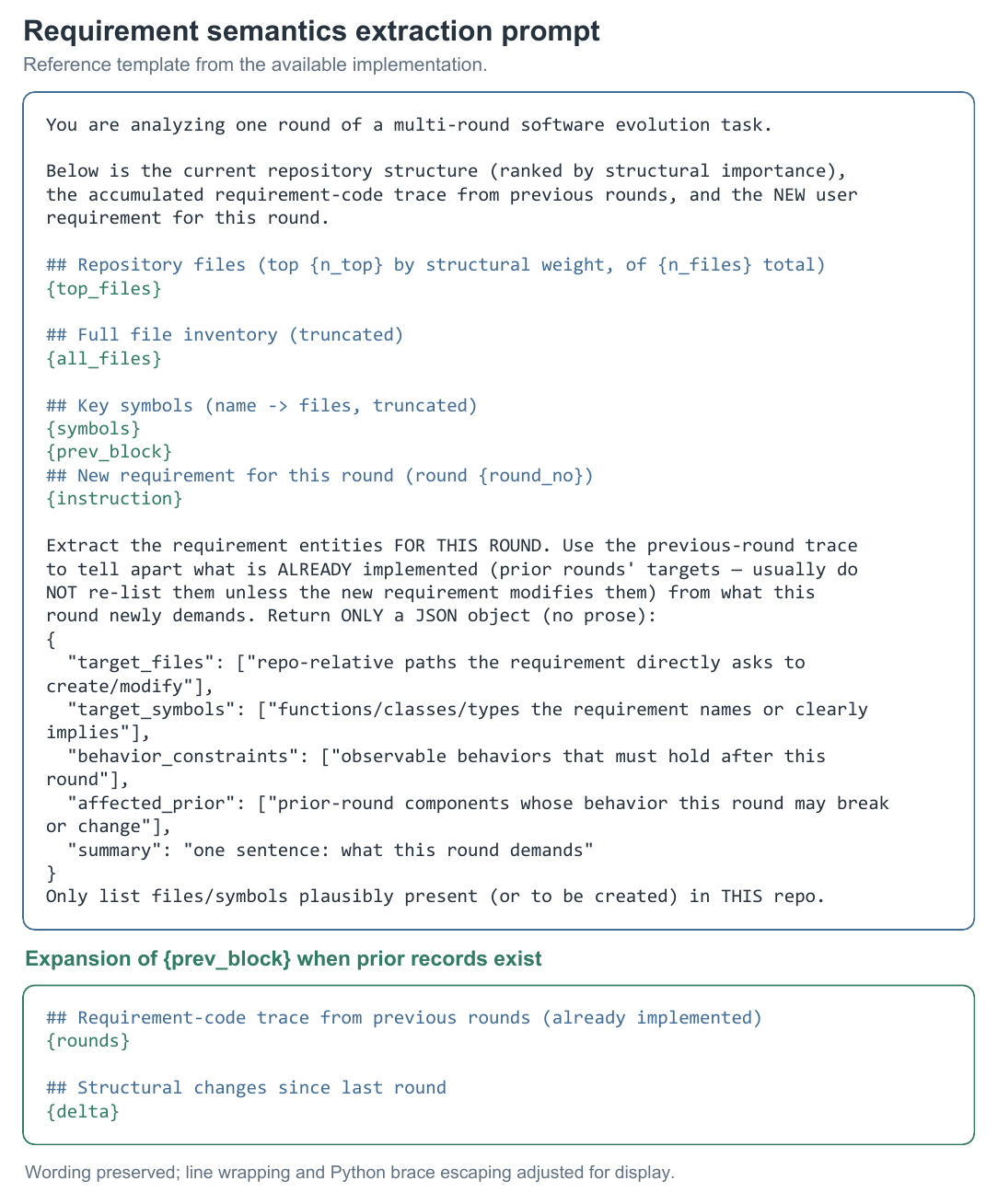}
    \caption{Reference prompt for requirement semantics extraction in Section~4.2. The upper box organizes the extraction inputs and specifies the fields of $Q_t$. The lower box expands \texttt{\{prev\_block\}} with prior requirement records and structural changes. The template wording is retained; line wrapping and Python brace escaping are adjusted for display.}
    \label{fig:repotrace_extraction_prompt}
\end{figure}

To identify code elements associated with $r_7$, the extraction step uses the inputs defined in Section~4.2: the current requirement, repository structure, structural changes, and prior requirement records. Figure~\ref{fig:repotrace_extraction_prompt} shows how the reference prompt organizes these inputs. The file lists and symbol-to-file associations summarize $G_t$; \texttt{\{instruction\}} supplies $r_t$; and \texttt{\{prev\_block\}} supplies $H_{<t}$ and $\Delta_t$ when prior records exist. The requested JSON fields correspond to the candidate target files and symbols, behavior constraints, potentially affected prior code elements, and summary in $Q_t$.

For JobForge Round~7, the saved extraction record identifies \texttt{critical.go} and \texttt{main.go} as candidate target files, together with symbols such as \texttt{criticalCmd} (Figure~\ref{fig:repotrace_structure_example}(b)). The behavior constraints specify the critical-path output and require all existing subcommands to remain unchanged. The \texttt{affected\_prior} field is empty: no specific prior code elements are listed in that field. The preservation requirement is therefore represented by a behavior constraint, rather than by an explicit list of affected prior code elements. These fields provide the requirement-specific information used in the next step.

\subsection{Agent Context Augmentation Across Rounds}
\label{app:repotrace_message}
\begin{figure}[!ht]
    \centering
    \includegraphics[width=0.82\linewidth]{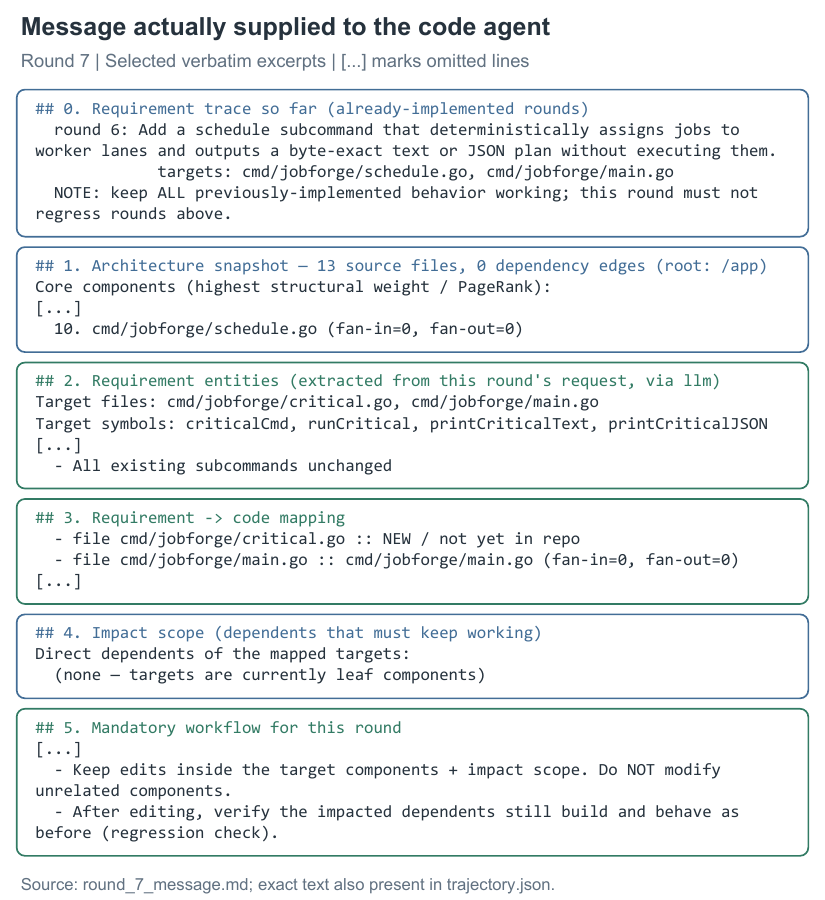}
    \caption{Excerpts from the requirement-code correspondence message $M_7$ supplied to the code agent. The message includes prior requirements, repository structure, candidate targets, existing/new-file mappings, and verification guidance. Ellipses mark omitted lines. The empty impact list reflects $E_7=\emptyset$; the original phrase \texttt{leaf components} does not establish the absence of source-code dependencies.}
    \label{fig:repotrace_message}
\end{figure}

The correspondence message connects the candidate targets in $Q_7$ to the repository described by $G_7$, illustrating the role of $M_t$ in Section~4.3. Figure~\ref{fig:repotrace_message} shows excerpts from the actual Round~7 message. Comparing the target paths with $V_7$ identifies \texttt{critical.go} as a candidate new file and \texttt{main.go} as an existing file. The message thus distinguishes creating the requested implementation from modifying existing code.

To retain the context of earlier requirements, the recorded message also includes the Round~6 scheduling requirement and its candidate targets, \texttt{schedule.go} and \texttt{main.go}. Together with the constraint that existing subcommands remain unchanged, this history reminds the agent to preserve scheduling behavior when adding \texttt{critical}. The verification guidance asks the agent to check that affected code still builds and preserves existing behavior. The complete message appears verbatim in the first user input of the Round~7 agent trajectory, alongside the current requirement.

This example shows how prior requirement records supplement the current requirement before code modification. As described in Section~4.3, RepoTrace retains graph snapshots and requirement records for subsequent rounds. These records preserve earlier analyses; they do not certify that the agent has completed a requirement or passed its tests.

\section{Setting for Requirement Focus Misalignment Analysis}
\label{app:misalignment_setting}

We systematically analyze 2,656 agent trajectories from EvoCode-Bench. Following the terminology used in the main text, each agent trajectory denotes the interaction trace associated with one model--task round. These trajectories cover 26 multi-round software evolution tasks and 12 code models.

For each trajectory at round $r$, we define $A_r$ as the set of source-code paths modified by the agent and $O_r$ as the set modified by the benchmark-provided oracle implementation. We statically extract $A_r$ from mutation operations recorded in the agent trajectory and extract $O_r$ from the corresponding \texttt{solve.sh} and solution artifacts. File paths are normalized relative to the repository root and deduplicated within each round. We retain source files written in Go, Python, JavaScript/TypeScript, Java, C/C++, Rust, Ruby, PHP, and shell. Tests, third-party dependencies, build outputs, caches, temporary files, and common documentation files are excluded using fixed path-based rules.

Before computing the aggregate metric, we apply a conservative parsing-quality filter. The static parser resolves concrete file paths for 19,431 of the 19,926 detected mutation-related command instances, corresponding to a parsing coverage of 97.52\%. We exclude any trajectory containing an unresolved mutation operation because its extracted modification set may be incomplete. This excludes 143 trajectories. One additional trajectory is excluded because its trajectory artifact is unavailable, leaving 2,512 trajectories with reliably extracted path sets.

The main analysis focuses on failed trajectories and excludes Round~1, which often involves repository initialization and has a qualitatively different modification scope from subsequent evolution rounds. This selection results in 1,846 trajectories. We measure file-level deviation from the oracle modification scope as
\begin{equation}
    \mathrm{OffOracle}
    =
    \frac{\sum_{r} |A_r \setminus O_r|}
         {\sum_{r} |A_r|}.
\end{equation}
Under this setting, 2,132 of the 3,865 agent-modified source-file instances fall outside the oracle modification scope, yielding an off-oracle ratio of 55.16\%.

This metric measures disagreement with the modification scope of the benchmark reference implementation; it does not assume that the oracle provides the only valid implementation. The analysis is path-based and does not distinguish content-equivalent, whitespace-only, or comment-only writes. It also cannot verify whether every statically detected command successfully modified the workspace. We therefore report 55.16\% as a corpus-level static estimate rather than a full-corpus replay ground truth. In a 30-trajectory calibration, the static and replay estimates under the same failed-trajectory, non-Round-1 setting are 32.26\% and 34.29\%, respectively.

\section{HTML Report of Repository Changes Across Rounds}
\label{app:repotrace_html}

We provide an HTML report of the JobForge trajectory in the Supplementary Material. The report organizes repository snapshots by round and displays candidate targets, extracted symbols, potentially affected prior code elements, and added files alongside file and function-call diagrams. Readers can browse the report in a web browser. Code agents equipped with file or browser tools can inspect the same HTML artifact. The supplied report is a static document that supports inspection of the recorded changes.

\begin{figure}[!ht]
    \centering
    \includegraphics[width=0.95\linewidth]{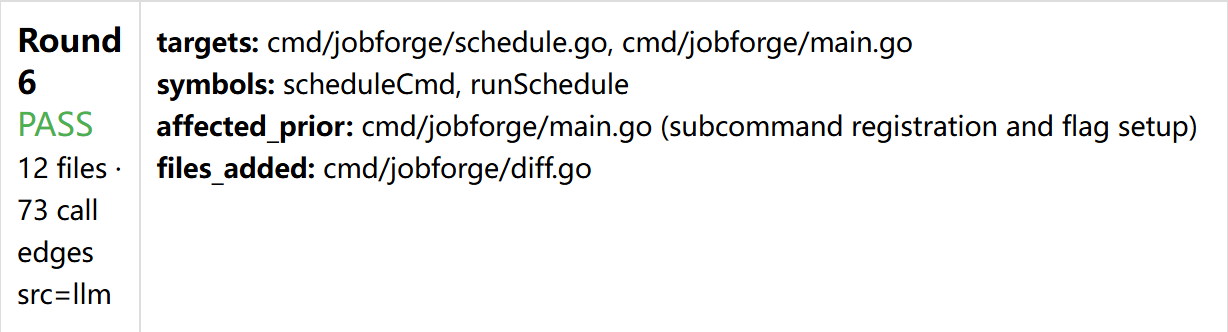}\\[2pt]
    {\small (a) Round 6: 12 files; \texttt{schedule.go} is a candidate target.}\\[7pt]
    \includegraphics[width=0.95\linewidth]{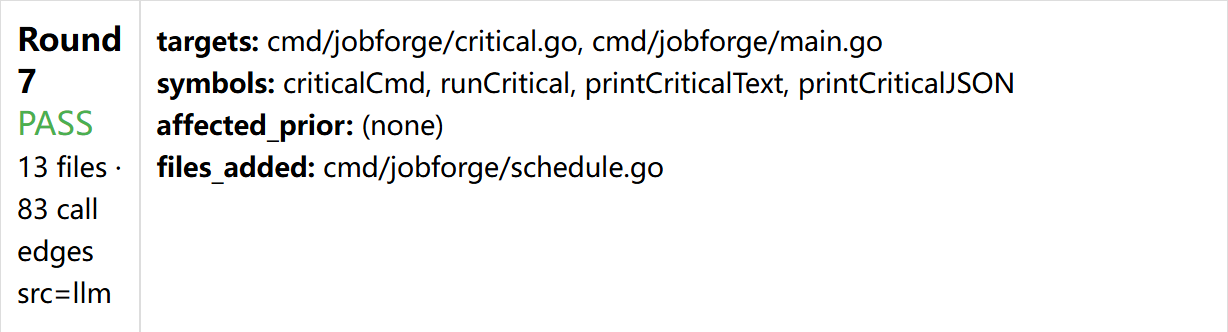}\\[2pt]
    {\small (b) Round 7: 13 files; \texttt{schedule.go} appears as an added file.}\\[7pt]
    \includegraphics[width=0.95\linewidth]{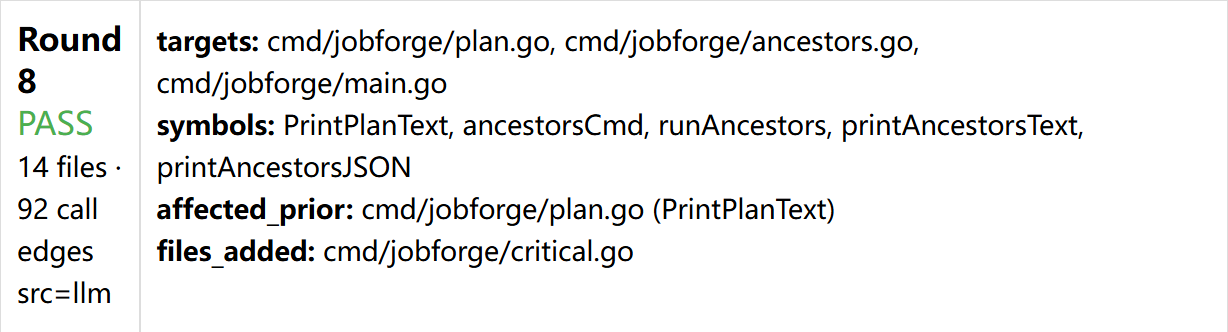}\\[2pt]
    {\small (c) Round 8: 14 files; \texttt{critical.go} appears as an added file.}
    \caption{Cropped screenshots of three consecutive records in the supplementary HTML report. The report combines extraction fields with function-call graphs re-collected from a reconstructed agent workspace. Its call-edge counts therefore describe the supplementary reconstruction, not the saved file-relation sets in Appendix~\ref{app:repotrace_structure}.}
    \label{fig:html_round_changes}
\end{figure}

Figure~\ref{fig:html_round_changes} shows excerpts from Rounds~6--8. The file count increases from 12 to 13 and then 14. In Round~6, \texttt{schedule.go} appears as a candidate target. In Round~7, \texttt{schedule.go} appears in the added-file record, and \texttt{critical.go} becomes a candidate target. In Round~8, \texttt{critical.go} appears in the added-file record. These entries distinguish the targets of a current requirement from file additions observed in a subsequent snapshot. See the Supplementary Material for the complete nine-round report.

\begin{figure}[!ht]
    \centering
    \includegraphics[width=0.95\linewidth]{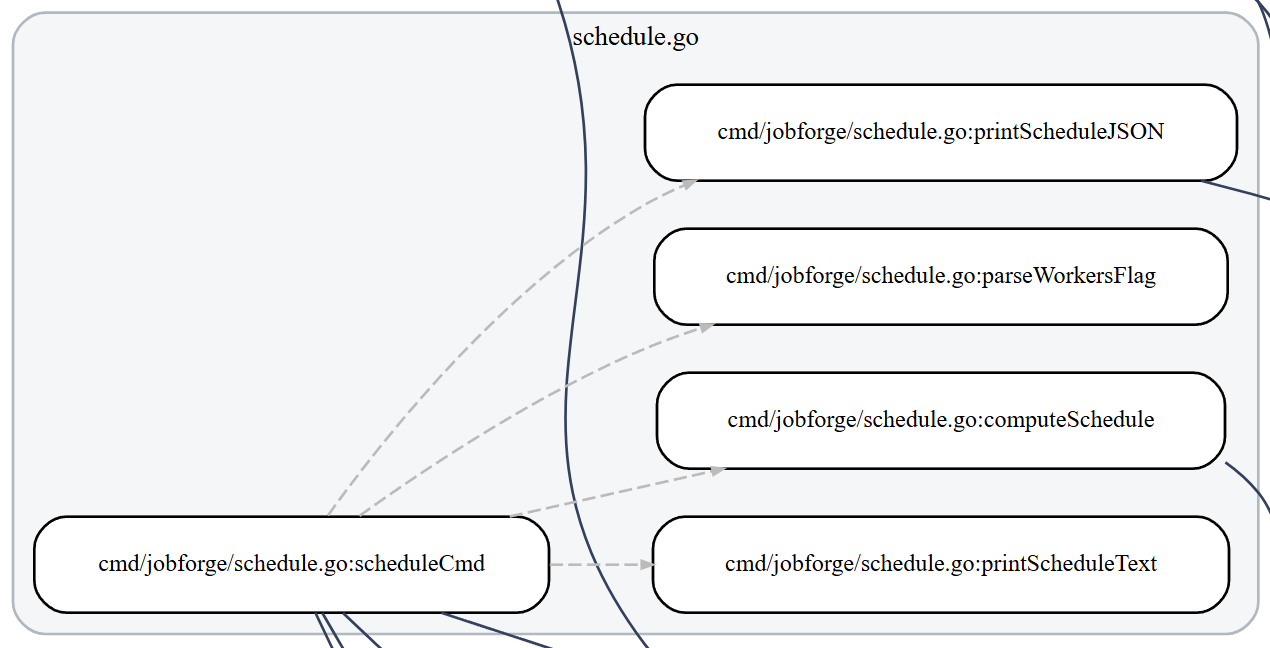}
    \caption{The \texttt{schedule.go} region of the Round~7 HTML diagram. The screenshot shows \texttt{scheduleCmd} and functions for parsing worker settings, computing the schedule, and formatting output. This file corresponds to the added-file entry in Figure~\ref{fig:html_round_changes}(b).}
    \label{fig:html_schedule_detail}
\end{figure}

\begin{figure}[!ht]
    \centering
    \includegraphics[width=0.95\linewidth]{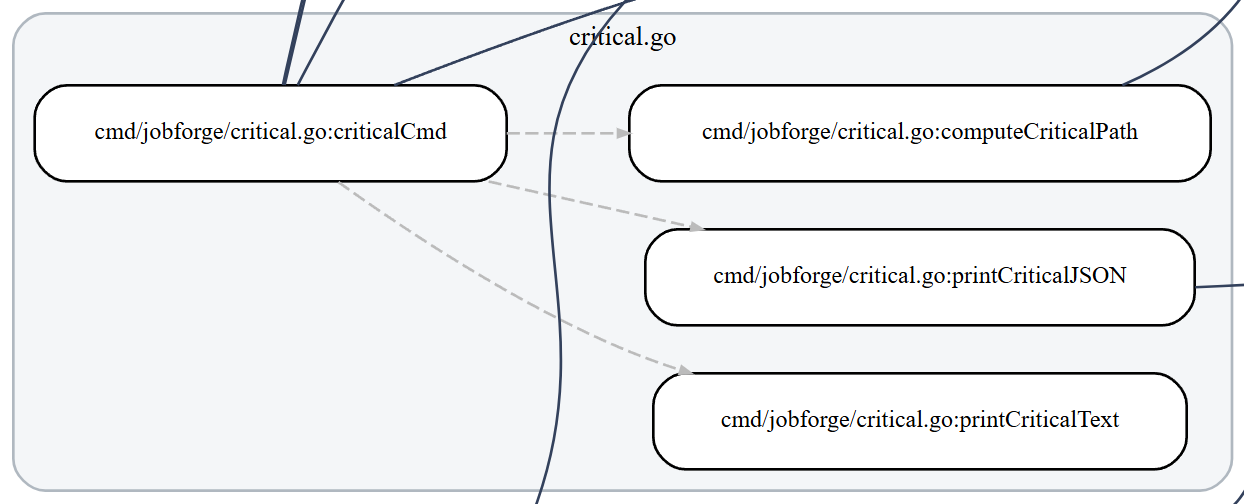}
    \caption{The \texttt{critical.go} region of the Round~8 HTML diagram. This file corresponds to the added-file entry in Figure~\ref{fig:html_round_changes}(c). The full diagram, including connections to other files, is available in the supplementary HTML report.}
    \label{fig:html_critical_detail}
\end{figure}

\clearpage
Figures~\ref{fig:html_schedule_detail} and~\ref{fig:html_critical_detail} show corresponding file-level details from the reconstructed function-call diagrams. The outer boxes group functions by file. Solid arrows denote cross-file calls, and dashed arrows denote calls within a file. The screenshots retain the original labels and show local regions of the larger diagrams; some cross-file arrows continue beyond the crop boundaries.

\begin{figure}[!ht]
    \centering
    \includegraphics[width=\linewidth]{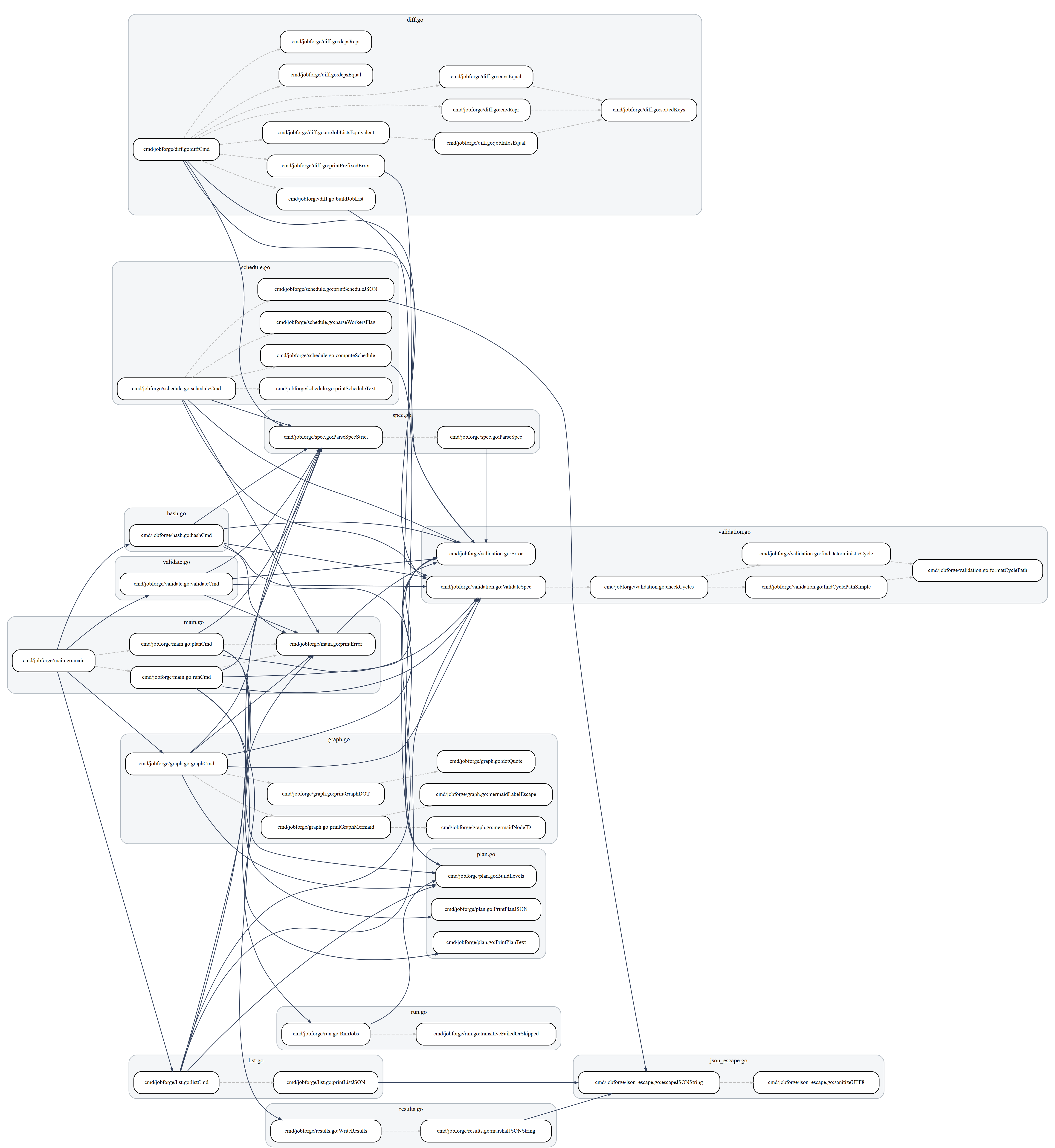}
    \caption{The overall repository structure and function-call graph for Round~7 in the supplementary HTML report.}
    \label{fig:html_round7_overall}
\end{figure}

\end{document}